\documentclass[twocolumn,prl,aps,epsfig,superscriptaddress]{revtex4} 
\usepackage{graphicx,subfigure} 
\usepackage{dcolumn} 
\usepackage{graphicx} 
\usepackage{bm} 
\usepackage{amssymb} 

\begin{document} 
 
\title{Thermodynamic behavior of a phase transition in a model 
for sympatric speciation} 
\author{K. Luz-Burgoa} 
\address{Instituto de F\'{\i}sica, Universidade Federal 
Fluminense, Campus da Praia Vermelha, Boa Viagem, Niter\'oi, 24210-340, RJ, 
Brazil} 
\email{karenluz@if.uff.br} 
\author{S. Moss de Oliveira} 
\address{Instituto de F\'{\i}sica, Universidade Federal 
Fluminense, Campus da Praia Vermelha, Boa Viagem, Niter\'oi, 24210-340, RJ, 
Brazil} 
\author{Veit Schw\"ammle} 
\address{Instituto de F\'{\i}sica, Universidade Federal 
Fluminense, Campus da Praia Vermelha, Boa Viagem, Niter\'oi, 24210-340, RJ, 
Brazil and Institute for Computational Physics
Pfaffenwaldring 27, D-70569 Stuttgart} 
\author{J. S. S\'a Martins} 
\address{Instituto de F\'{\i}sica, Universidade Federal 
Fluminense, Campus da Praia Vermelha, Boa Viagem, Niter\'oi, 24210-340, RJ, 
Brazil} 
 
\date{\today} 
 
\begin{abstract} 

We investigate the macroscopic effects of the ingredients
that drive the origin of species through sympatric speciation. In our
model, sympatric speciation is obtained as we tune up the strength of
competition between individuals with different phenotypes. As a function
of this control parameter, we can characterize, through the behavior of a
macroscopic order parameter, a phase transition from a non-speciation to a
speciation state of the system. The behavior of the first derivative of
the order parameter with respect to the control parameter is consistent
with a phase transition and exhibits a sharp peak at the transition point.
For different resources distribution, the transition point is shifted, an
effect similar to pressure in PVT system.  The inverse of the parameter
related to sexual selection strength behaves like an external field in the
system and, as thus, is also a control parameter.  The macroscopic effects
of the biological parameters used in our model reveal thus fingerprints
typical of thermodynamic quantities in a phase transition of an
equilibrium physical system.

\end{abstract} 

\maketitle 
 
The branching of a single population into two or more species without
prevention of gene flow through geographic segregation is known as
sympatric speciation \cite{editor,mayr,bush,futuyma}. Herbivorous insects
have long been considered prime candidates for sympatric speciation
because of an intimate and frequently highly specialized relationship with
their host plants, which serve as habitat, food resource, and, often,
mating location \cite{mallet}.  The apple maggot fly {\it Rhagoletis
pomonella} has been considered, since 1966, as the classical example of
sympatric speciation in progress \cite{bush}. {\it R. pomonella} shifted
from feeding on the unabscised fruit of its native host hawthorn ({\it
Crataegus spp.}) to utilizing the introduced, domesticated apple ({\it
Malus pumila}) sometime in the mid-1800s in the Hudson River Valley region
of the state of New York.  Genetic evidence suggests that the species is
in the process of shifting and adapting to this new host plant
\cite{feder1,feder2}.

Two ingredients are important for sympatric speciation to happen in 
a population \cite{sara,van,turelli}: 
The competition caused by fluctuations in ecology \cite{kk,dd} 
and assortative mating caused by selective mating \cite{takimoto,gavrilets}. 
Ecological and sexual selection models have addressed these two aspects of 
sympatric speciation separately \cite{van}. 
The starting point of ecological models is the assumption that sympatric 
speciation results 
from disruptive selection. That is, competition for diverse resources leads to 
separation in a population, if individuals with intermediate phenotypes are  
losers when they compete with those with extreme ones. 
Such selection can cause sympatric speciation because 
it provides an advantage for reproductive isolation between opposite,  
well-adapted, extreme phenotypes, and reproductive isolation can be achieved 
due to evolution of nonrandom mating \cite{kksexo}. 
Sympatric speciation can also be driven by selective mating, or sexual 
selection, that is, 
nonrandom mating leading to differential reproductive successes of different 
phenotypes. For example if the choice of a mate depends on two traits: 
male display 
(e.g. nuptial hue, varying from red to blue through purple) and female 
preference for variants of display. Some females may prefer red males and 
others prefer blue males, this can tear the population apart and create a 
pair of species consisting of red-prefering females and red males and of 
blue-prefering females and blue males. 

To study sympatric speciation by simulations we 
use the individual-based Penna model \cite{penna}. 
In previous work with this model \cite{medeiros,ksjsabjp33,sjpkatcsbook} 
an abrupt ecological change was the drive that provoked disruptive selection, 
as in \cite{kk}, which led to 
speciation through the development of assortative mating. 
A different strategy was used to simulate sympatric speciation of 
predators in a food web \cite{foodwebPRE}.
In this case, three types of intra-specific competition were adopted, 
depending on the phenotypic group of the predators, and their 
strength was kept constant during the 
whole simulation.
In particular, a parameter $X$ was introduced, establishing the 
fraction of the populations of extreme phenotypic predators with which the 
intermediate phenotypic individuals would compete, besides competing among 
themselves. 
In the present paper we adopt the same kind of constant intra-specific 
competition and study first when speciation is achieved, 
depending on the value of $X$, for a uniform resource distribution 
per phenotype and a sexual selection of constant strength. 
We show that the competition strength $X$ plays the role of a control 
parameter in a phase transition, and that the fraction of sexual  
selective females in the population shows behavior similar to an order 
parameter. Furthermore, we show that the transition point and its functional 
form depend on the carrying capacity distributions and sexual selection 
strengths.

In the present model, competition for food and assortative mating are related 
to the same phenotypic trait. This trait is represented by a new pair of non 
age-structured bit-strings, which are crossed and recombined in the 
breeding process \cite{foodwebPRE}. 
The phenotypic characteristic is 
measured by counting, in this new pair of bit-strings, the number of bit 
positions where both bits are set to 1, plus the number of dominant positions 
(chosen as $16$) with at least one of the two bits set. It will therefore 
be a number $k$ between $0$ and $32$, which we will refer to as 
the individual's phenotype. 
We fix the mutation probability per locus, $0 \rightleftarrows 1$, 
of this phenotypic trait at $0.01$.  

In order to consider intra-specific competition depending on the individual's 
phenotype $k$, we modified the logistic Verhulst factor introducing three 
intra-specific competition terms, each one related to a given phenotypic 
group: 
\begin{equation}\label{fatorall} 
V(k,t) = \left\{\begin{array}{rcl} V_1(k,t), & 0\le k<n_1;&{\rm especialist}\\ 
V_m(k,t), & n_1\le k\le n_2;&{\rm intermediate}.\\ 
V_2(k,t), & n_2< k\le 32;&{\rm especialist}. 
\end{array}\right. 
\end{equation} 
As in the original Penna model, at every time step, and for each individual, 
a random real number uniformly distributed between $0$ and $1$ is generated; 
if this number is smaller than $V(k,t)$, the individual dies. 
For the extreme phenotype groups the competition is given by: 
\begin{equation}\label{extreme} 
V_{1(2)}(k,t) = \frac{P_{1(2)}(k,t)+P_m(k,t)}{F}, 
\end{equation} 
where $P_{1(2)}(k,t)$ accounts for the population with phenotype $k<n_1$ 
($k>n_2$) at time $t$, 
respectively, $P_m(k,t)$ accounts for the population with phenotype $k\in 
[n_1,n_2]$, and 
$F$ is a constant proportional to the carrying capacity, taken as 
$2\times 10^5$ in our simulations. 
Individuals with intermediate phenotypes ($P_m$) compete among themselves and 
also with a fraction $X$ of each population presenting an extreme phenotype.
The Verhulst factor for them is: 
\begin{equation}\label{intermediate} 
V_m(k,t) = \frac{P_m(k,t)+X*\left[P_1(k,t)+P_2(k,t)\right]}{F}, 
\end{equation} 
where $X$ can be thought of as the strength of competition between 
intermediate and extreme phenotypic populations. 
Eq.\ref{extreme} means that individuals with extreme phenotypes  
($P_1$, $P_2$) compete with those belonging to the same phenotypic group 
and also with the whole intermediate population, 
but there is no competition between extreme phenotypes of different groups 
because we are assuming they are specialized to some extent 
($[0,n_1=13)$,$(n_2=19,32]$) on particular resources.
 
In order to consider assortative mating, 
we introduce into each female genome a single locus (bit) that codes for this 
selectiveness, 
also obeying the general rules of the 
Penna model for genetic heritage and mutation. If it is set to $0$, the female 
is not selective in mating (panmictic mating). It is selective 
(assortative mating) if this locus is set to $1$. 
The mutation probability for this locus, which can be in both 
directions ($0\rightleftarrows 1$), is $0.001$.  
Mutated females that are born selective 
choose mating partners according to the following mating strategy: 
If a female has phenotype $k<16$ ($k>16$), it chooses, 
among $N_m$ males, the one with the smallest (largest) phenotype value $k$; 
If a selective female has $k=16$ then it chooses randomly to act as one of the 
above. 
Notice that with this strategy all females reproduce every time step 
from age $R=10$ until death. 

\begin{figure}[htbp] 
\begin{center} 
\subfigure{ 
\includegraphics[width=3.8cm,height=3.4cm]{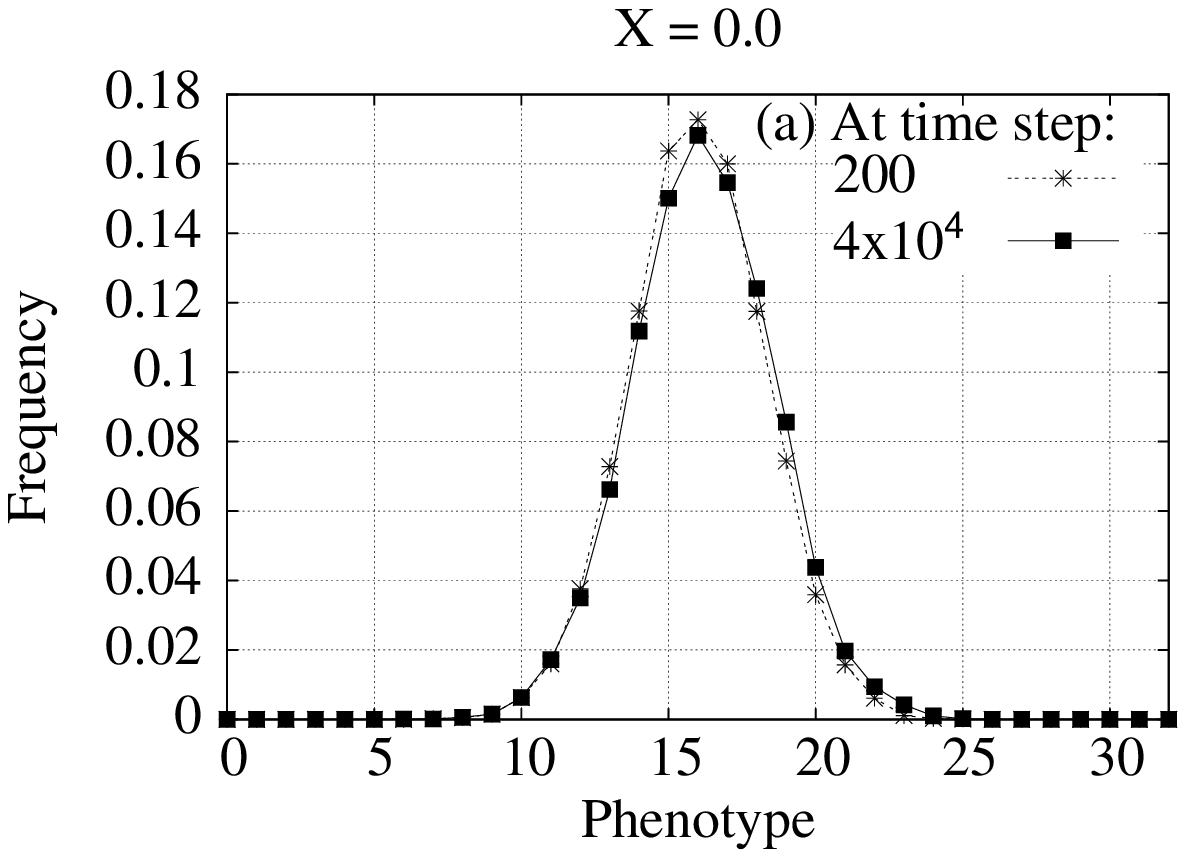}}\subfigure{ 
\includegraphics[width=3.8cm,height=3.4cm]{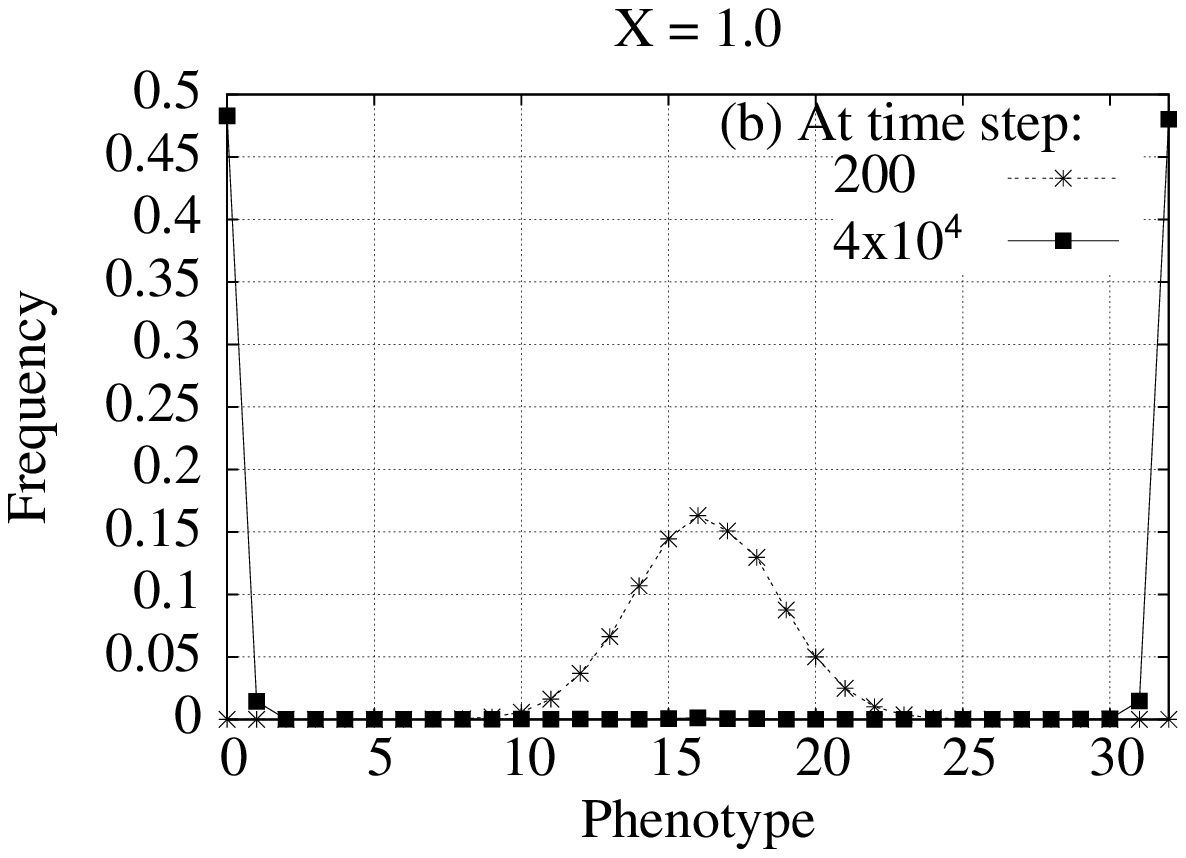}}\\ 
\subfigure{ 
\includegraphics[width=3.8cm,height=3.4cm]{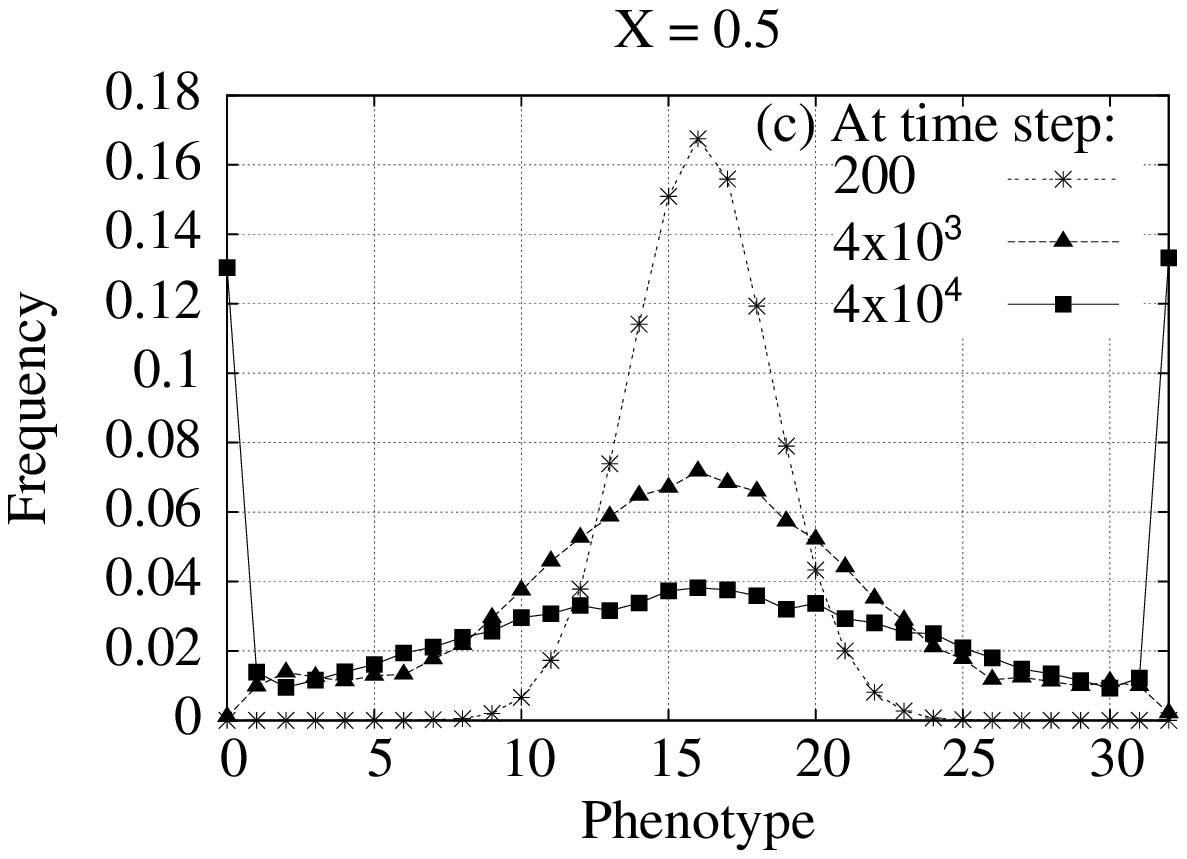}}\subfigure{ 
\includegraphics[width=3.8cm,height=3.4cm]{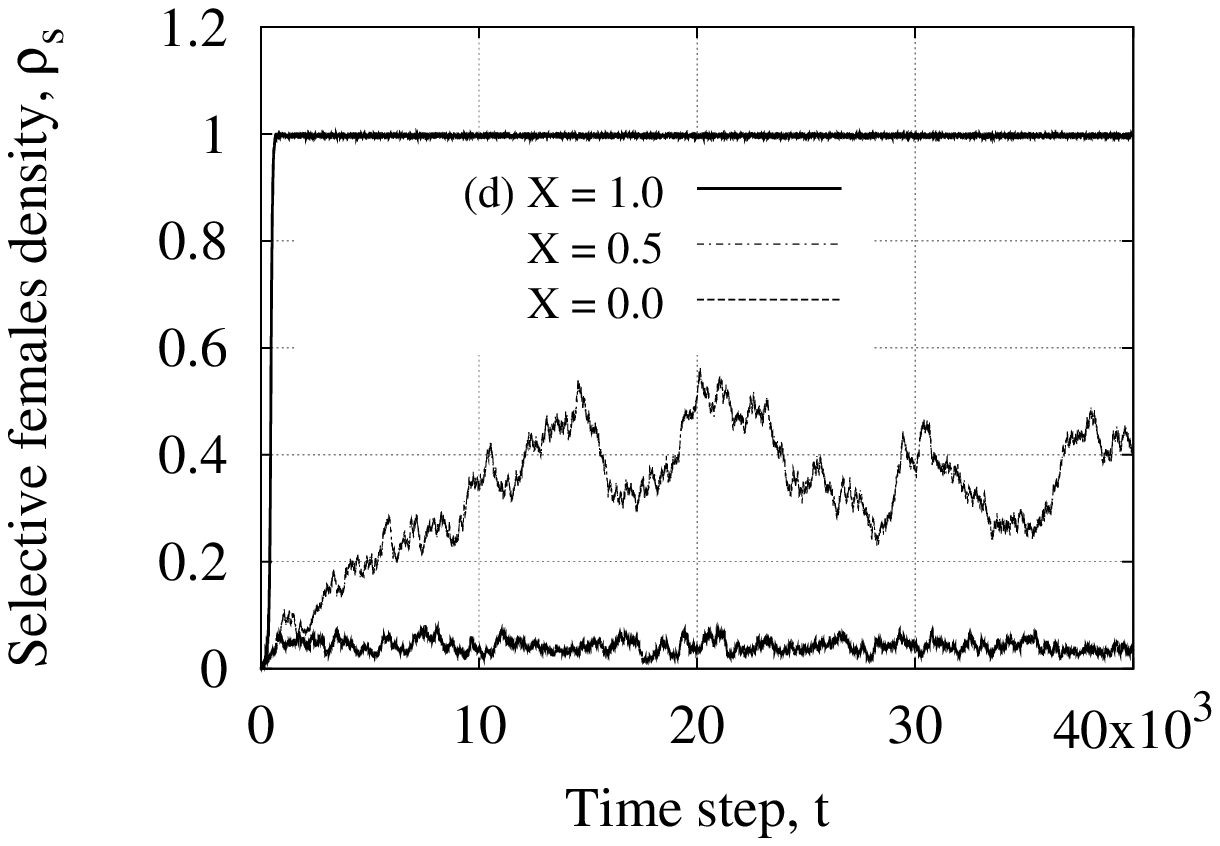}}
\end{center} 
\caption{In (a), (b) and (c) the phenotype distributions of the whole 
population for different $X$. 
In the initial steps of the simulation, $t=200$, the distribution is, 
in all cases, a gaussian 
centered at intermediate phenotypes. 
For (a) and (b), the distributions at $t=4\times10^3$ are
equal to those at $t=4\times10^4$ and are stationary.
For (c) the distribution is not stationary neither at 
$4\times 10^3$ nor at $4\times10^4$ (see text). 
(d) The time behavior of the selective females density. 
} 
\label{figsympatric} 
\end{figure} 

At the beginning of the simulations females are non-selective and all the 
$6\times10^3$ individuals (half males and half females) have a random 
phenotype. 
For the parameter $X=0$, the population with intermediate phenotypes 
does not compete with the extreme phenotypic ones [Eq. \ref{intermediate}]
and, in fact, suffer less competition than the 
other two [Eq. \ref{extreme}]. In this case, nearly all females remain 
non-selective, see Fig.\ref{figsympatric}(d) lower line, and  
the phenotype distribution corresponds 
to a stationary gaussian function centered at $k=16$,
Fig.\ref{figsympatric}(a) squares.
As opposed to the situation for $X=0$, when we introduce a strong competition 
for the population with intermediate phenotypes, by setting, say, $X=1.0$, 
only the individuals with extreme phenotypes $k=0$ and $k=32$ survive, as 
shown in Fig.\ref{figsympatric}(b) squares. 
In this case the density of selective females goes to $\rho_s\approx 1$ very 
fast, that is, the females with extreme phenotypes mate only with males of 
its same phenotypic group, see Fig.\ref{figsympatric}(d) upper line. 
This means that there are two new 
sympatric species, reproductively isolated.
For the competition strength $X=0.5$, the phenotype distribution is not 
stationary: in runs that differ by the choice of the seed of the random 
number generator, the final distribution sometimes has one maximum at  
$k=16$, Fig.\ref{figsympatric}(c) triangles, and some other times  
it has two maxima at $k=0$ and $k=32$, Fig.\ref{figsympatric}(c) squares. 
The time behavior of the density of selective females presents large 
fluctuations, see Fig.\ref{figsympatric}(d) central line. 
Fig. \ref{figsympatric}d shows an important change in the population 
organization, from a non-speciation state with $\rho_s\approx 0$, to a 
sympatric speciation state with $\rho_s\approx 1$, as we increase the 
strength of competition, $X$, for the intermediate phenotypes. To 
determine the range of values of $X$ for which sympatric speciation may be 
obtained, we will analyze the behavior of the mean density of selective 
females, for many different strengths of competition.

\begin{figure}[htbp] 
\begin{center} 
\subfigure{ 
\includegraphics[width=4.2cm,height=3.5cm,angle=0]{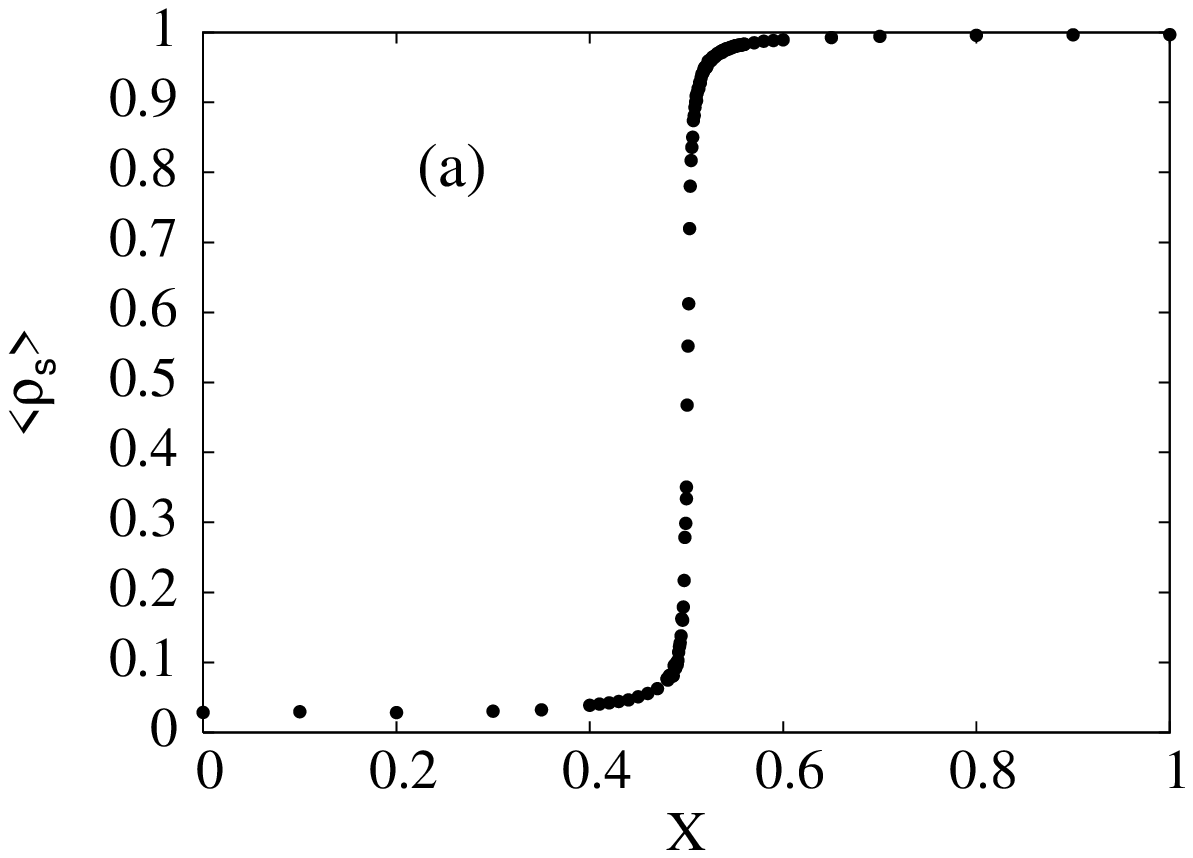}}\subfigure{ 
\includegraphics[width=4.2cm,height=3.5cm,angle=0]{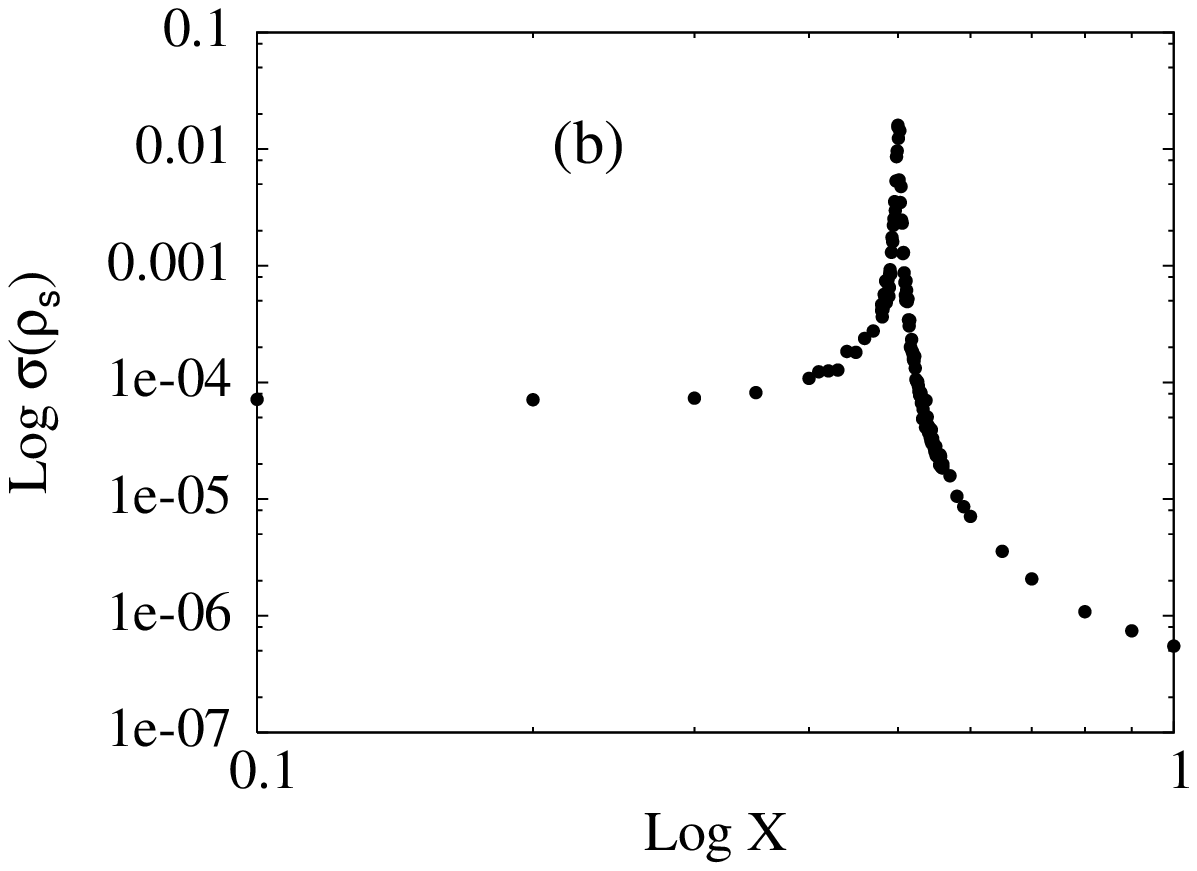}}
\end{center} 
\caption{a) Mean values the selective females density 
the order parameter of the speciation transition, as function of the control 
parameter $X$. 
b) Logarithm scale of the standard deviation versus $X$.
For each value of $X$ we have made 10 simulations with the same parameters, 
but using different initial seeds for the random number generator. 
In each simulation we calculate the mean value of the density of selective 
females during the last $10^4$ time steps, and then average the results of 
the ten runs. 
}
\label{figtran} 
\end{figure} 
 
The behavior of the mean density $<\rho_s>$ as a function of $X$ is shown
in Fig.\ref{figtran}(a). The population changes rather abruptly from a
non-speciation to a speciation state when we change slightly the strength
of competition, close to $X_c=0.5$ \cite{tese}.
Another fingerprint of the macroscopic effect, $X$ on $<\rho_s>$,
is the peak shown by the logarithm of
the first derivative of the order parameter at $X_c$,
Fig.\ref{figtran}(b).  
These behaviors are very
similar to what happens to an order parameter as a function of the control
parameter in an equilibrium phase transition of a physical system.
This transition separates a single-species phase from
one in which two species coexist in sympatry. 
In the single-species phase, Fig. \ref{figtran}(a) $X<X_c$,
the population presents a high diversity with many different phenotypes in
the population, see Fig.\ref{figsympatric}(a) squares,
and has a mean size of $\approx 25\times10^3$. 
In the two-species phase, Fig. \ref{figtran}(a) $X>X_c$,
the mean size of the whole population is $\approx 50\times10^3$, or twice
the value of the former phase, and the phenotypes in the population
cluster around only two distinctively separated values, 
Fig. \ref{figsympatric}(b) squares.  
In the Fig. \ref{figtran}(b), the large values attained
by $\sigma(\rho_s)$ just above $X_c$ 
arise from large
fluctuations in the number of individuals.


Ecological conditions have been considered an essential ingredient for 
divergence and speciation in sympatry \cite{matthew}. To evaluate its 
importance in a phase transition context, we simulated different ecological 
conditions by modifying the carrying capacity of the environment, which has 
so far been considered as a constant $F$ in the Eqs. 
\ref{extreme} and \ref{intermediate}. 
It will now be phenotype-dependent and will drive 
the population to experience a disruptive selection between the specialist 
and intermediate phenotypes. Its general functional form is: 
$F_{\sigma_k}=2\times10^5*e^{-(k-16)^2/\sigma_k^2}$,
where each individual, with phenotype $k$, will feed on a different resource 
niche: For small values of $\sigma_k^2$, 
the specialists will have fewer resources than 
individuals with intermediate phenotype. 
\begin{figure}[htbp] 
\begin{center} 
\includegraphics[width=6.0cm,height=3.8cm]{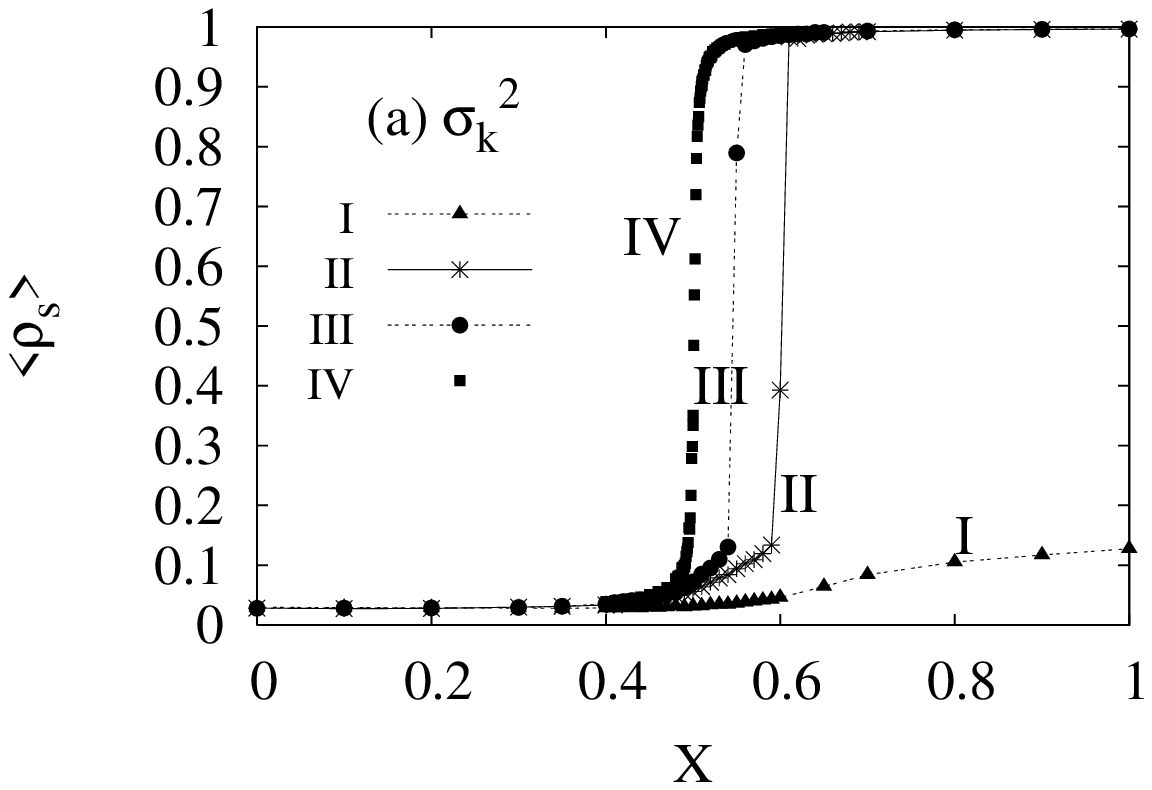}
\includegraphics[width=6.0cm,height=3.8cm]{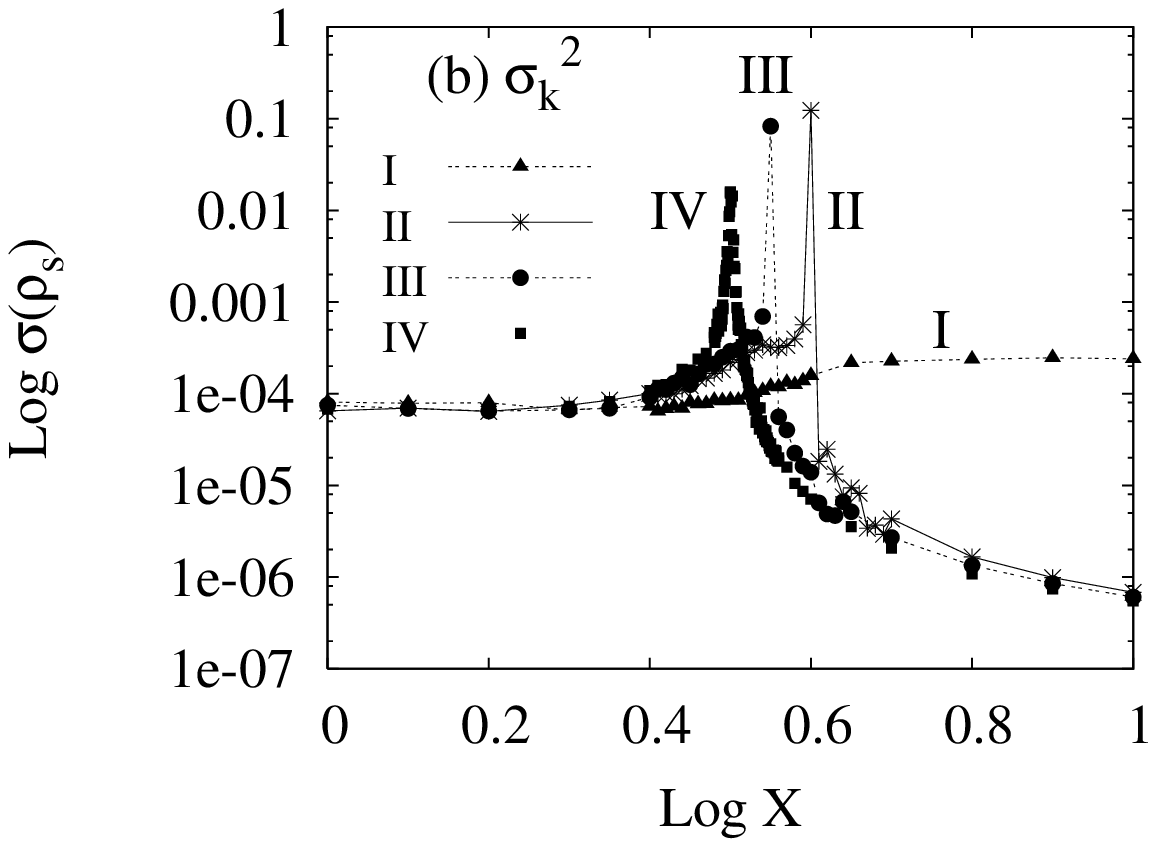}
\end{center} 
\caption{The figures show the effect of disruptive selection
on the speciation transition. The carrying capacity from I to 
III correspond to $F_{\sigma_k}$ with $\sigma_k^2 = 
10^3,5\times10^3,10^4$, respectively. $F_{\sigma_k}=F$ 
correspond to IV. } 
\label{figpres} 
\end{figure} 

In Fig. \ref{figpres}(a) the macroscopic effect of the carrying capacity 
is the shift suffered by the transition point for different values of 
$\sigma_k$, which has then an effect similar to pressure in PVT systems. 
For small value of $\sigma_k^2$ and for $X>X_c$, case I in Figs. 
\ref{figpres}(a) and (b), the population 
prefers a non-speciation state, even in the presence of a high  
competition for intermediate phenotypes. This happens because there are not 
enough resources for two groups with extreme phenotypes. It is nevertheless 
important to notice that the population has a large diversity in this 
case. That is, the phenotype distribution looks like 
Fig. \ref{figsympatric}(c) triangles, but it is a stable 
distribution, see Fig. \ref{figpres}(b) $I$. 

\begin{figure}[htbp] 
\begin{center} 
\subfigure{ 
\includegraphics[width=3.9cm,height=3.2cm]{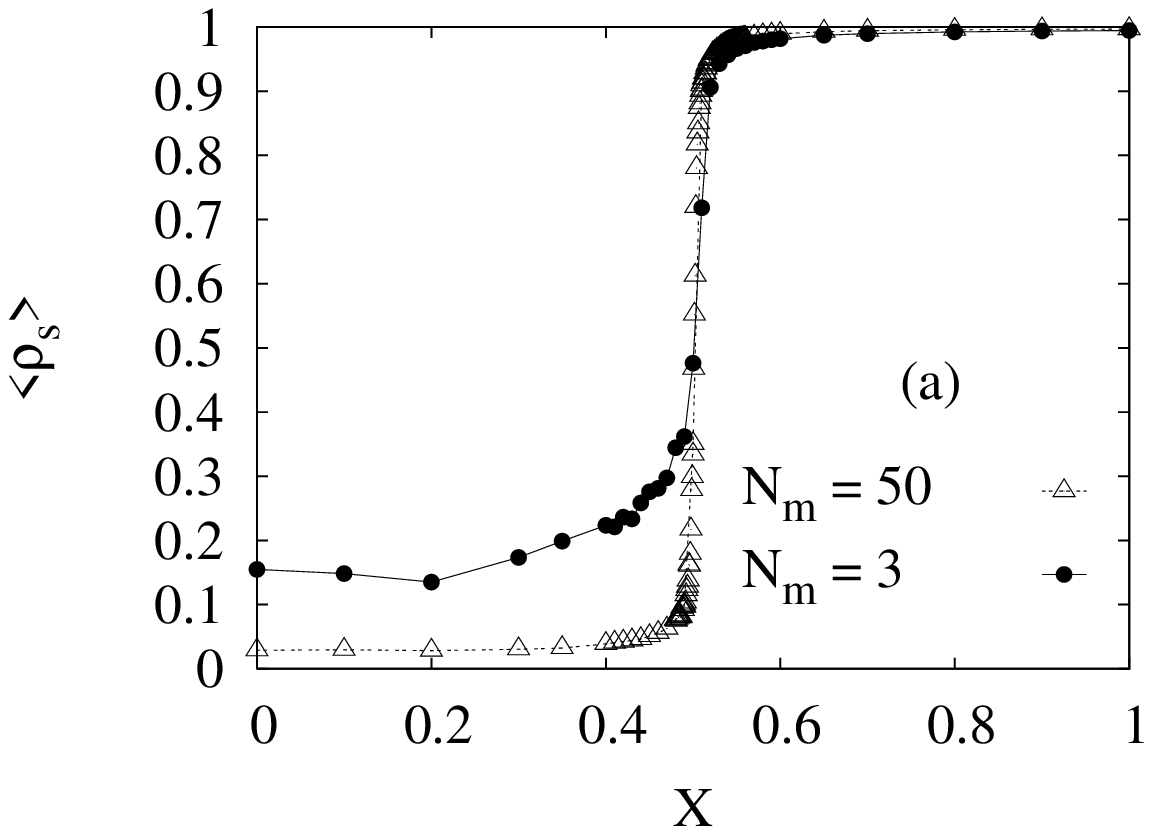}}\subfigure{ 
\includegraphics[width=3.9cm,height=3.2cm]{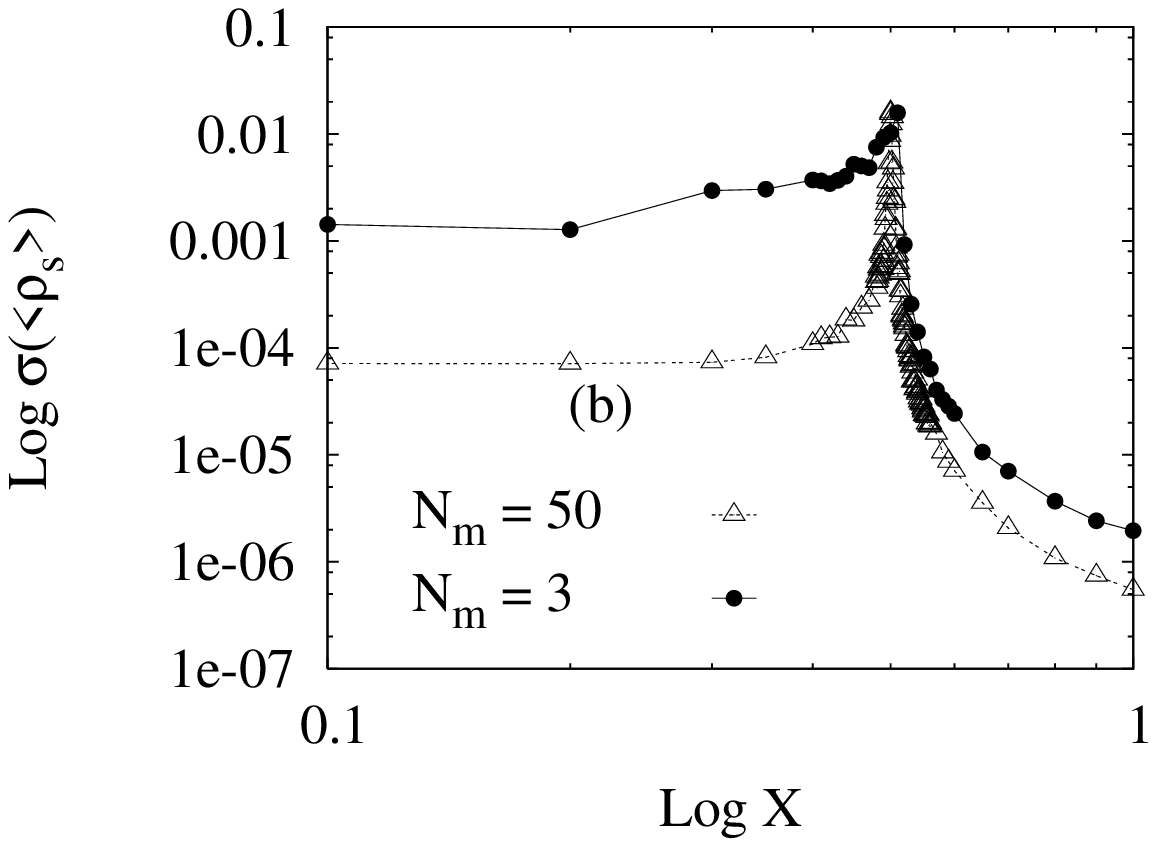}}
\end{center} 
\caption{In (a) and (b) the effect of the sexual 
selection strength, 
$N_m$. It is important to point out that the 
simulation time for $N_m=3$ was $8\times10^5$, $20$ times bigger than in 
the other cases.} 
\label{figcampo} 
\end{figure} 

Sexual selection in the population is associated to the number of mating 
choices each female performs before reproduction, the parameter $N_m$. 
The probability of a selective female with phenotype 
$k<16$  to mate with a male of opposite phenotype is 
$P_{<16}\approx(0.5)^{N_m}$. In the previous section, $N_m = 50$ and 
this probability is almost zero, meaning that the selective females are 
highly discriminatory against the opposite phenotype. With $N_m = 3$ the 
probability becomes $P_{<16}=0.125$ and it is then  possible for a selective 
female with $k<16$ to mate with a male of phenotype $k>16$. 
When we reduce the number of mating choices per female, we can see that 
the phase transition is destroyed, Figs. \ref{figcampo}(a) and (b). 
For an equilibrium physical system the phase transition 
disappears when there is an applied 
external field, as, for an example, happens to the paramagnetic transition 
of magnetic materials at the Curie point.
A small value for sexual selection strength, equivalent to 
the application of a magnetic field, produces an increase 
of the selective female density,
see Fig. \ref{figcampo}(a) circles for $X<X_c$. 
This density approaches $0.5$ since the difference between 
selective mating and panmictic mating is small. 

In conclusion, we reported an investigation of the macroscopic effects 
of the parameters ($X$ competition, $F_{\sigma_k}$ disruptive
natural selection and $N_m$ sexual selection strength) on 
the origin of species by sympatric speciation, in a model 
where fitness and mate choice are represented 
by the same trait.
For this model and in the context of the phase 
transition, Fig. \ref{figtran}, it was possible 
to quantify the ingredients that promote speciation 
in sympatry. 
The macroscopic effects of the parameters related to natural selection 
by resources are in qualitative agreement with other individual-based  
models that studied the necessary ecological conditions for sympatric 
speciation \cite{sara,dd,kk}. 
The effect of sexual selection strength behaves like an external
field in the system and in an physical system a field is a control 
parameter like pressure or temperature.
Another characteristic of selective 
mating is related to the relaxation time since it depends on 
the $N_m$ value, see caption in Figs. \ref{figcampo}a and b.  
These results, for assortative mating are in qualitative agreement 
with \cite{takimoto,gavrilets}.

We believe these analogies between biological and physical 
parameters will help in the understanding of the sympatric 
speciation process.

\begin{acknowledgments} 
We would like to thank to T.J.P. Penna, J. Nogales, 
J. F. Stilck, E. Curado and P.M.C. de Oliveira for fruitful discussions. 
We acknowledge financial support from the agencies CLAF, CNPq, and CAPES. 
\end{acknowledgments} 
 

\end{document}